# Circular displacement current induced anomalous magneto-optical effects in high index Mie resonators

Shuang Xia[1,2], Daria Ignatyeva[3,4,5], Qing Liu[6], Hanbin Wang[7], Weihao Yang[1,2], Jun Qin*[1,2], Yiqin Chen[6], Huigao Duan*[6], Yi Luo[7], Ondřej Novák[8], Martin Veis*[8], Longjiang Deng[1,2], Vladimir I. Belotelov*[3,4,5] and Lei Bi*[1,2]

[1]National Engineering Research Center of Electromagnetic Radiation Control Materials, University of Electronic Science and Technology of China, Chengdu 610054, China

[2]State Key Laboratory of Electronic Thin-Films and Integrated Devices, University of Electronic Science and Technology of China, Chengdu, 610054, China

[3]Lomonosov Moscow State University, Leninskie Gory, Moscow 119991, Russia

[4]Vernadsky Crimean Federal University, *295007,* Simferopol, Russia

[5]Russian Quantum Center, *143025,* Moscow, Russia

[6]College of Mechanical and Vehicle Engineering National Engineering Research Center for High Efficiency Grinding, Hunan University Changsha, 410082, China

[7]Microsystem and Terahertz Research Center, China Academy of Engineering Physics, Chengdu 610200, China

[8]Charles University of Prague, Faculty of Mathematics and Physics, Ke Karlovu 3, 12116 Prague 2, Czech Republic

*Email: qinjun@uestc.edu.cn, duanhg@hnu.edu.cn, v.belotelov@rqc.ru, veis@karlov.mff.cuni.cz, bilei@uestc.edu.cn

**ABSTRACT**: Dielectric Mie nanoresonators showing strong light-matter interaction at the nanoscale may enable new functionality in photonic devices. Recently, strong magneto-optical effects have been observed in magneto-optical nanophotonic devices due to the electromagnetic field localization. However, most reports so far have been focused on the enhancement of conventional magneto-optical effects. Here, we report the observation of circular displacement current induced anomalous magneto-optical effects in high-index-contrast Si/Ce:YIG/YIG/SiO$_2$ Mie resonators. In particular, giant modulation of light intensity in transverse magnetic configuration up to 6.4 % under *s-polarized* incidence appears, which is non-existent in planar magneto-optical thin films. Apart from that, we observe a large rotation of transmitted light polarization in the longitudinal magnetic configuration under near normal incidence conditions, which is two orders of magnitude higher than for a planar magneto-optical thin film. These phenomena are essentially originated from the unique circular displacement current when exciting the magnetic resonance modes in the Mie resonators, which changes the incident electric field direction locally. Our work indicates an uncharted territory of light polarization control based on the complex modal profiles in all-dielectric magneto-optical Mie resonators and metasurfaces, which opens the door for versatile control of light propagation by magnetization for a variety of applications in vectoral magnetic field and biosensing, free space non-reciprocal photonic devices, magneto-optical imaging and optomagnetic memories.

Dielectric Mie nanoresonators showing strong light-matter interaction at the subwavelength scale have attracted great research interest recently[1-4]. Compared to plasmonic devices in which metal nanostructures are used to confine electromagnetic fields at the nanoscale, all-dielectric Mie resonators show several important differences, including lower optical absorption loss, much larger field penetration into the dielectric nanostructures, and the existence of unique magnetic resonance modes[5, 6]. These features make dielectric Mie resonators and metasurfaces a fertile playground to discover novel photonic phenomena. Observations such as magnetic mirror[7, 8], directional scattering[9, 10], enhanced optical nonlinear effects[11, 12] and photoluminescence[13] have been demonstrated, which are promising for future nanophotonic device applications.

Recently, enhancement of the magneto-optical effects has been reported in dielectric photonic nanostructures. Several theoretical works predict strong Faraday rotation enhancement in all-dielectric magneto-optical nanostructures[14, 15], as well as large Magneto-optical Kerr Effect (MOKE) enhancement in all-dielectric gratings.[16] Experimentally, all-dielectric magneto-optical metasurfaces featuring quasi-waveguide modes have been fabricated[17, 18]. Transverse magneto-optical intensity effect is observed upon the excitation of waveguide modes[19, 20]. Meanwhile, silicon Mie resonators are reported to enhance the Faraday rotation and transverse magneto-optical intensity effect in a 5 nm thick metallic nickel film[21, 22]. Electric dipole resonance induced circular birefringence and dichroism are observed in a perpendicular magnetic anisotropic Pt/Co film[23]. However, a high index contrast, all-dielectric magnetic metasurface exhibiting strong Mie resonance modes have not been considered in this respect. How the Mie resonance modes influence the magneto-optical effects remains largely unexplored.

Here, we observe the effect of the circular displacement current in the high index contrast Mie resonators on the magneto-optical activity. The effect is disclosed on the structure of Si nanodisks on a magneto-optical bilayer with $Y_3Fe_5O_{12}$(YIG) and $Ce_1Y_2Fe_5O_{12}$ (Ce:YIG) films deposited on a quartz substrate (Fig. 1a). In the near infrared range Ce:YIG has a prominent magneto-optical activity much larger than that for YIG, therefore in this case most of the magneto-optical response comes from Ce:YIG. Due to the large index contrast, the metasurface exhibits strong Mie resonance modes including the magnetic dipole (MD) mode, electric dipole (ED) mode, electric quadrupole (EQ) mode, magnetic quadrupole (MQ) mode, as well as waveguide (WG) modes. The highly confined electromagnetic field in Ce:YIG leads to strong light-matter interaction and circular displacement currents which induce anomalous magneto-optical effects in transversal and longitudinal magnetic configurations. It should be emphasized that these effects are absent or negligible in planar Ce:YIG thin films. These results indicate promising potential of controlling light propagation by utilizing the complex Mie resonance modes in all-dielectric magneto-optical Mie resonators and metasurfaces. Our work may inspire the development of novel magneto-nanophotonic devices such as vectoral magnetic field sensing, free space non-reciprocal photonic devices, magneto-optical imaging and optomagnetic memories.

## RESULTS AND DISCUSSION

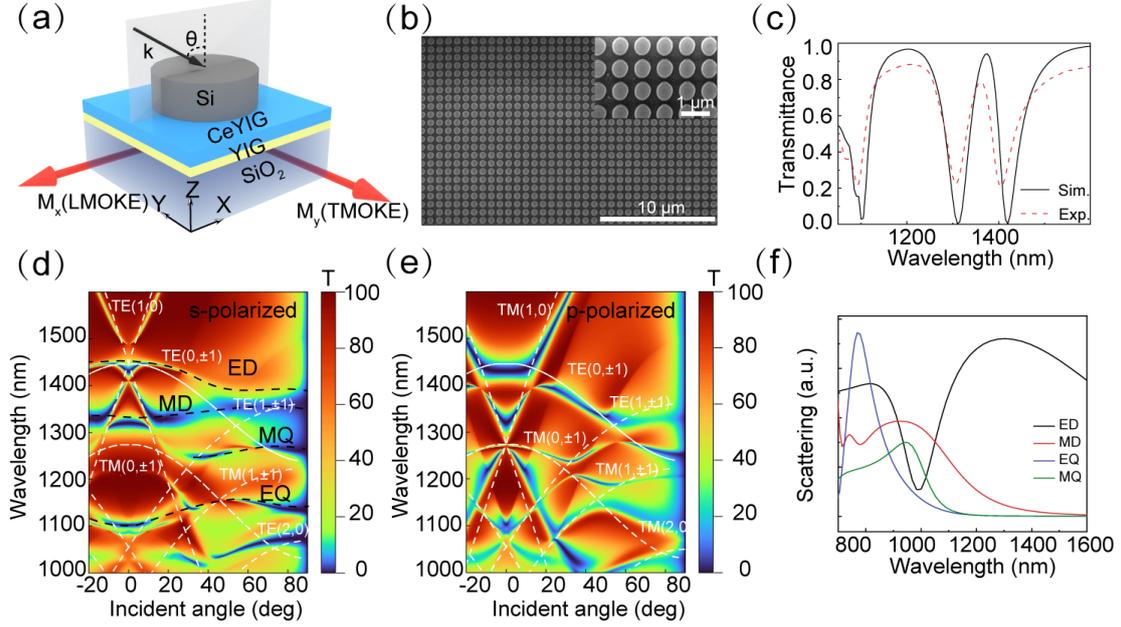

**Figure 1. Device structure and mode analysis of the all-dielectric magneto-optical metasurface** (a) Schematic diagram of the metasurface magnetized in Voigt configurations for s-polarized TMOKE and LMOKE-T measurements. The device consists of periodic silicon nanodisks on Ce:YIG/YIG bilayer thin films fabricated on a quartz substrate. (b) Top-view SEM image of a fabricated sample. The inset shows a zoom-in image of several Si nanodisks. (c) Measured and simulated transmission spectra of the device under normal incidence and linear polarization. (d)(e) Simulated transmission spectra of the array with R=230 nm cylinder radius as a function of the wavelength and incident angles for s-polarized and p-polarized incidence respectively using rigorous coupled wave analysis (RCWA) method. The white dash lines indicate the dispersion of the WG modes in the metasurface calculated using the planar waveguide theory. (f) Analytical multipole analysis of scattering spectra for an individual silicon nanodisk (height 170 nm, radius 230 nm) embedded into a low-index medium (n=1.46).

The device consists of periodic Si ($n$=3.17) nanodisk Mie resonators on Ce:YIG ($n$=2.30) /YIG ($n$=2.30) bilayers fabricated on double side polished $SiO_2$ ($n$=1.46) substrates (Fig. 1a). The YIG and Ce:YIG thin films were fabricated by pulsed laser deposition (PLD) and annealing (see Methods). The thicknesses of the YIG and Ce:YIG layers were 50 nm and 200 nm, respectively. The X-ray diffraction and Faraday rotation characterizations are shown in Supplementary Fig. S1. The Si nanodisks were fabricated by plasma enhanced chemical vapor deposition (PECVD) of amorphous silicon thin films followed by electron beam lithography (EBL) and reactive ion etching (RIE) (see details in Supplementary Fig. S2 and Methods). The period, radius and thickness of the Si pillars are P=750 nm, R=230 nm and H=170 nm, respectively (Fig. 1b). In order to study the magneto-optical effects in transversal (TMOKE) and longitudinal (LMOKE-T) configurations, we magnetized the device along y- and x- directions, respectively (Fig. 1a).

We firstly simulated the transmission spectrum of the sample using finite element method (COMSOL MUITIPHYSICS) (see details in Supplementary Fig. S3). According to the directions of the applied magnetic field, the permittivity tensor of Ce:YIG and YIG takes the form:[24, 25]

$$\varepsilon_{MO} = \begin{bmatrix} \varepsilon_{xx} & aM_z & aM_y \\ -aM_z & \varepsilon_{yy} & -aM_x \\ -aM_y & aM_x & \varepsilon_{zz} \end{bmatrix} \quad (1)$$

where $\varepsilon_{ii}$ and $aM_i$ (*i* stands for x, y or z) represent the diagonal and off diagonal permittivity tensor elements, respectively. Fig. 1c shows the transmittance spectrum of the magneto-optical Mie resonators under normal incidence with linear polarized light along x-direction. Three resonance peaks at wavelengths of 1100 nm, 1310 nm and 1420 nm are clearly seen. The experimental transmittance spectrum (red dotted line in Figure 1c) is in a very good agreement with the simulation results (see measurement set-up and details in Methods). The small difference of transmission intensity and resonant wavelengths between the experiment and simulation may be attributed to the sample imperfections during the fabrication process, including slight size deviations, non-vertical sidewalls *etc*. Simulated transmission spectra as a function of wavelengths and incident angles for s-polarized and p-polarized incident light (Fig. 1d, e, respectively) demonstrate different dispersive behavior of the resonances. To identify their character we calculated the dispersion relation of the waveguide modes according to the device geometry and using the planar waveguide theory[18, 19, 26] (see white dashed lines in Fig. 1d, e). These modes correspond to different diffraction orders of the Si nanodisk grating as labeled in the figures (see Methods). Several modes with narrow linewidth agree well with the calculated waveguide modes. However, other modes cannot be explained by waveguide modes, for example the modes (transmittance dips) at high incident angles labeled by black dashed lines for s-polarized incidence, whose wavelengths seem to be independent on the incident angle, and the full width at half maximum (FWHM) is quite broad. These modes can be attributed to different Mie resonance modes as indicated by multipole analysis results for normal incidence as shown in Fig. 1f (see Methods). The scattering cross-section is computed on the basis of a multipole decomposition by considering an individual silicon nanodisk embedded into a low-index medium ($n$=1.46)[9]. The multipole analysis agrees with experimental observations despite of a blue shift of the Mie resonance wavelengths, which is due to the more complicated surrounding medium environment including multilayer YIG/CeYIG films, also the coupling effect between different nanodisks. For s-polarized incidence, the Mie resonance wavelengths are almost angle independent, as shown by black dashed lined in Fig. 1d[27]. As the incident angle increases, ED and MD resonances show a tendency to couple with each other, and the MQ resonance becomes increasingly apparent[28]. Characteristic circular electric field and displacement current distribution are observed both in Si and Ce:YIG for MD and MQ resonances, as shown in Supplementary Fig. S5. For the p-polarized incidence, EQ and ED resonance wavelengths are almost independent on the incident angles for small incident angles. The mode at 1310 nm for the non-zero angles of incidence splits into two separate modes[27]. This dispersion relationship is consistent with the TM (1,0) WG modes, where (1,0) indicate the grating diffraction order[29-32]. Therefore, in the case of small incident angles, the Mie resonance and these WG modes couple with each other, which also explains why the resonance peaks of the modes in p-polarization is broader than that in s-polarization for small incident angles[33]. As the angle of incidence increases, the contribution of the WG modes dominates and the width of the resonance peak gradually narrows. The near-field distributions of these modes with s- and p- polarization under different incident angles are shown in details in Supplementary (Figs. S5 and S6).

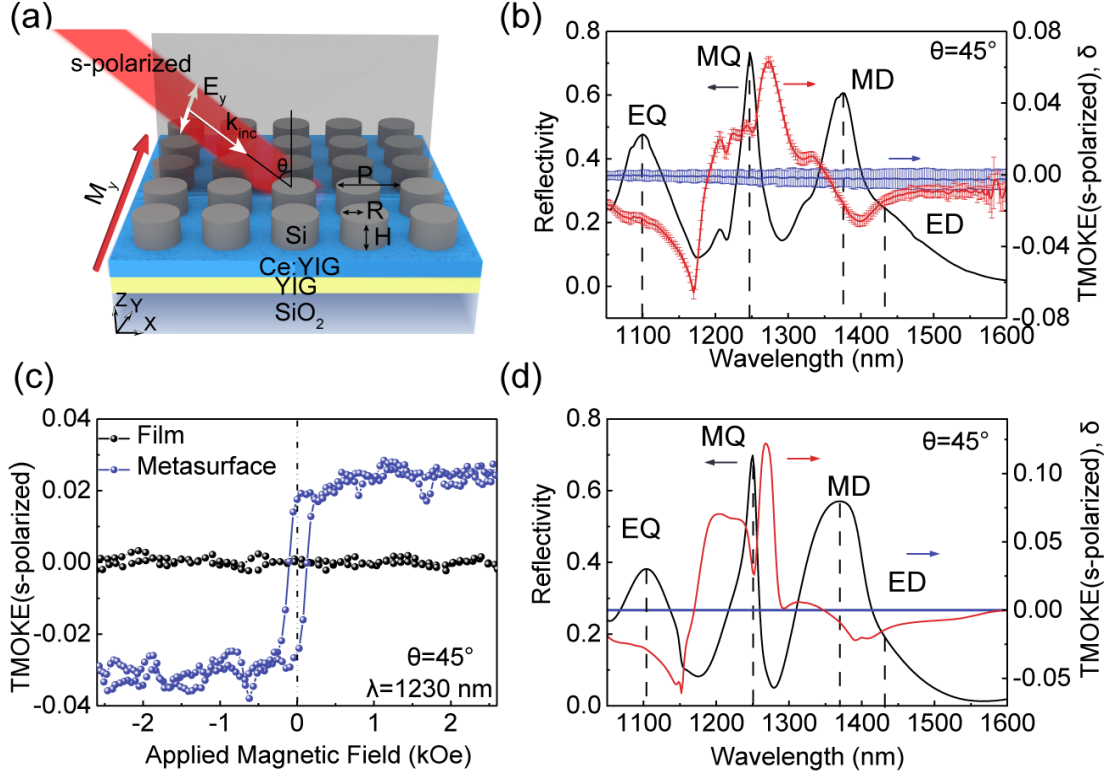

**Figure 2. Reflection spectrum and giant TMOKE for s-polarized incident light** (a) Schematic diagram of s-polarized TMOKE characterization set-up. All measurements are obtained for s-polarized light and θ=45° incidence angle. (b) Measured s-polarized TMOKE and reflection spectra of the metasurface compared with bare Ce:YIG/YIG thin films. The error bars are the standard deviation of 5 consecutive measurements. (c) Measured TMOKE hysteresis for the metasurface and the bare MO films at 1230 nm wavelength. (d) Simulated reflection spectra and TMOKE responses of the metasurface for s-polarized light and θ=45° incidence angle.

Next, we characterize the transverse magneto-optical Kerr effect (TMOKE) spectra for s-polarized light incident at $\theta=45°$. The incidence plane is parallel to the (1,0) direction of the Si nanodisk grating. The schematic diagram of s-polarized TMOKE measurement is shown in Fig. 5a. The value of TMOKE is defined as the relative change in reflectivity due to the magnetization switch along the incidence plane normal direction[34, 35]:

$$\delta = 2\frac{R(+M) - R(-M)}{R(+M) + R(-M)} \quad (2)$$

According to Fig. 1d, the reflection spectrum in Fig. 2b (black curve) shows high reflectivity at 1100 nm, 1250 nm and 1370 nm corresponding to the EQ, MQ and MD modes, respectively. And a relatively weak resonance peak at 1420 nm wavelength is attributed to the ED mode. The electric and magnetic near-field distribution corresponding to these resonances are detailed in

Supplementary Fig. S5.

In the planar magneto-optical thin films, it is well known that there is no TMOKE under s-polarized incident light[36, 37], as confirmed by the blue line in Fig. 2b measured on bare Ce:YIG/YIG thin films, which is zero within the measurement error. However, a large s-polarized TMOKE is observed in the magneto-optical Mie resonators, as shown by the red line in Fig. 2b. Importantly, we observe high reflectivity up to 73 % together with strong s-polarized TMOKE up to $\delta$=2.7 % at the MQ mode wavelength of 1250 nm. Stronger s-polarized TMOKE up to $\delta$=6.4 % at 1275 nm wavelength and $\delta$=-6.4 % at 1170 nm wavelength are also observed around the MQ resonances, which are attributed to low reflectance (~10 %, smaller value of the denominator in equation (3)) induced enhancement of the TMOKE, namely the optical contribution[38]. A TMOKE peak of $\delta$=-1.4 % also appears at 1375 nm wavelength with 61 % reflectivity, corresponding to the MD mode. Note a sign change of the TMOKE is observed for both MQ and MD modes, which is caused by a sign change of the numerator in Eq. (3). Meanwhile, the s-polarized TMOKE is also non-zero, but weaker at EQ and ED resonances. These measurement results agree very well with simulation using COMSOL as shown in Fig. 2d (see details in Methods).

To confirm our observation, we also measured the TMOKE hysteresis of the structure and film using a custom-built magneto-optical characterization set up (See Supplementary Fig. S8), as shown in Fig. 2c. A clear TMOKE hysteresis is observed for the metasurface sample with up to 3 % intensity variation at 1230 nm, which is in drastic contrast with the thin film sample which showed no hysteresis at all. The hysteresis resembles the hysteresis of the CeYIG/YIG films with the in-plane magnetization.

It should be noted that these s-polarized TMOKE values are even higher than recently reported conventional p-polarized TMOKE in all-dielectric magneto-optical gratings[18, 19, 36]. We also measured and simulated the TMOKE spectrum under conventional p-polarized incidence for the thin film and metasurface as discussed in Supplementary Fig. S7. Giant p-polarized TMOKE up to ~20 % is also observed, which are attributed to waveguide mode induced TMOKE enhancement, as also discussed in previous publications.[19]

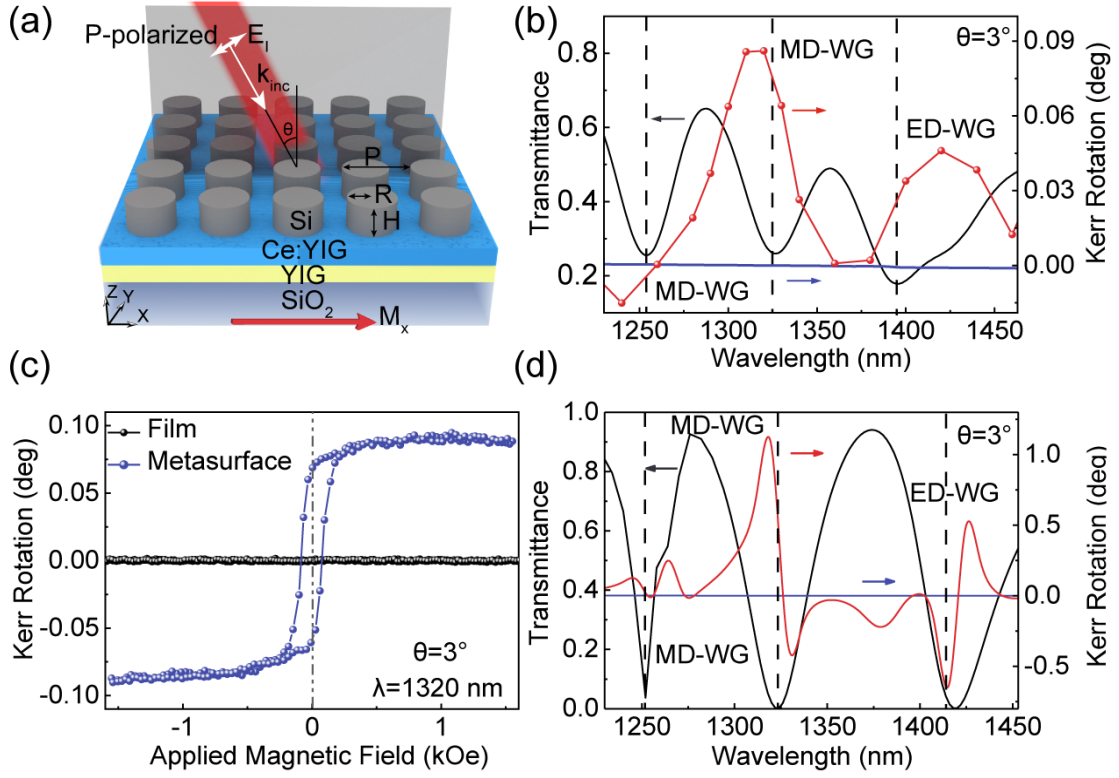

**Figure 3. Transmission spectra and giant LMOKE-T under p-polarzed incidence.** (a) Schematic diagram of LMOKE-T characterization set-up. All measurements are obtained for p-polarized light and under θ=3° incidence angle. (b) Measured LMOKE-T (red line with dots) and transmission spectra (black line) of the fabricated metasurfaces compared with bare Ce:YIG thin films (blue line). (c) Measured LMOKE-T hysteresis loops for the metasurface with R=230 nm and the bare MO film at 1320 nm wavelength. (d) Simulated transmission and LMOKE-T spectra of the metasurface.

Next, we study the LMOKE effect in transmission mode with sample configuration shown in Fig. 3a. The complex LMOKE-T angles $\phi_L$ can be expressed as[38]:

$$\phi_L = \psi + i\varphi = \arctan(\frac{t_{ps}}{t_{pp}}) \qquad (3)$$

where $\psi$ and $\varphi$ are the LMOKE-T angle and ellipticity, respectively. And $t_{ps}$ and $t_{pp}$ represent Fresnel transmission coefficients for s-polarized and p-polarized light respectively for p-polarized incidence. For symmetry considerations, a non-trivial LMOKE-T can only be observed when $[k \times N] \neq 0$[36], where k is the incident wave vector and N is the sample surface normal vector. Therefore, we measured the LMOKE-T and corresponding transmission spectra of the sample under θ=3°, a small off-normal incident angle, as shown in Fig. 3a. Fig. 3b shows the transmission and LMOKE-T spectra. Two waveguide modes are observed at 1250 nm and 1320 nm respectively as indicated by Fig. 1e. As discussed in Fig. 1e, the MD mode hybridizes with the WG mode at a small incident angle of 3°. For such a small angle, the LMOKE-T of the bare Ce:YIG/YIG thin film is almost negligible (<$10^{-3}$ deg). This value is smaller than our measurement noise, therefore it is only numerically simulated and shown by the blue line in Fig. 3b. Interestingly, the LMOKE-T shows a giant enhancement when exciting the resonance modes. A large LMOKE-T angle up to $\psi$=0.086 deg is observed at 1320 nm wavelength, which is about two orders higher compared to bare Ce:YIG/YIG films with same thicknesses of $\psi=9\times10^{-4}$ deg. Enhancement of LMOKE-T is also observed at 1250 nm and 1418 nm wavelength but with a lower amplitude, corresponding to the other waveguide mode and hybridized ED-WG mode respectively. This result can be better observed by comparing the LMOKE-T hysteresis loops between the metasurface and the film as shown in Fig. 3c. A clear hysteresis resembling the in-plane magnetization hysteresis of the Ce:YIG thin film is observed. The LMOKE-T angle of 0.086 deg is even comparable to the Faraday rotation angle of a bare film at the same wavelength as shown in Supplementary Fig. S1b. The simulated LMOKE-T spectrum is displayed in Fig. 3d, which shows similar characteristics with experiment results, despite of sharper peaks and larger rotation angle values. This is because in the case of simulation, the transmittance at resonances is almost zero, leading to a much larger optical contribution. The difference of transmission intensity between experiment and simulation may be caused by sample imperfections originated from the fabrication process.

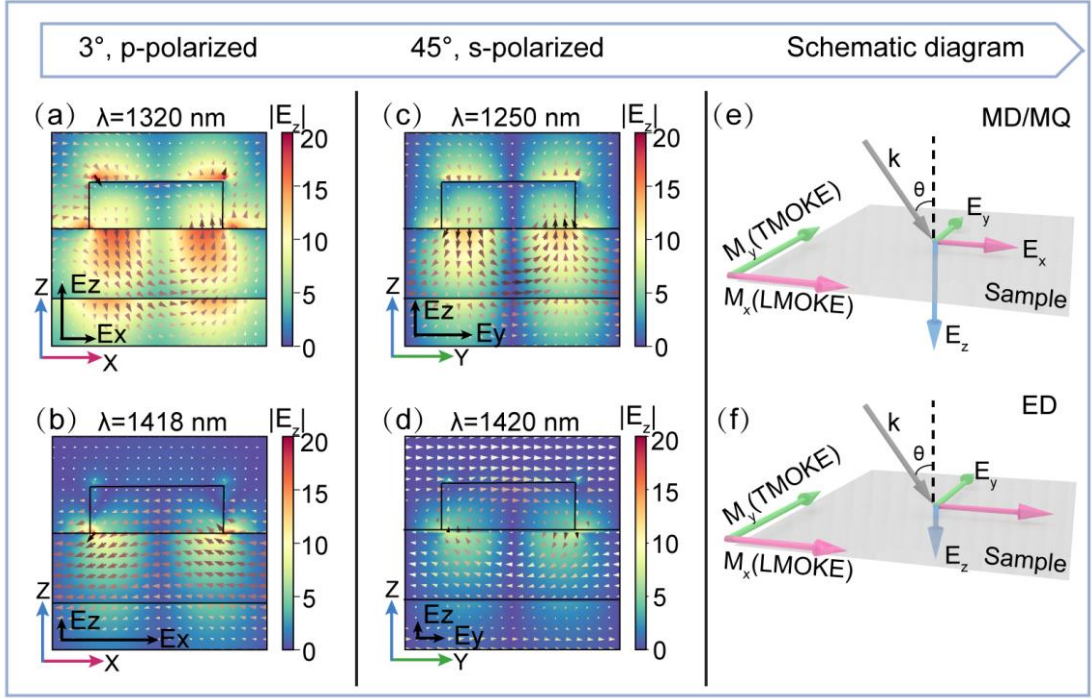

**Figure 4. Mechanism of the anomalous MO effects in the MO Mie resonators.** Electric field distribution for (a) MD resonance (b) ED resonance for θ=3°, p-polarized incidence (LMOKE-T case) and (c) MQ resonance (d) ED resonance for θ=45°, s-polarized incidence (TMOKE case) respectively. The pink arrows indicate the electric field vectors in Z-X plane. The vectors of different lengths at the bottom left corner of each figure indicate the relative magnitudes of the volume integral of $E_x$, $E_y$ and $E_z$ components in the magneto-optical layers. Schematic diagram of the relationship between electric field vectors, magnetic field direction and propagation direction are shown for (e) MD/MQ and (f) ED modes respectively.

To understand the mechanism of the observed magneto-optical effects, we consider the electric and magnetic near-field distribution for different resonance modes as shown in Fig. 4. Fig. 4a and 4b show the electric field profile under p-polarized incidence at θ=3° for the MD-WG and ED-WG mode wavelengths respectively (the LMOKE-T case). Fig. 4c and 4d show the modal profile under s-polarized incidence at θ=45° for the MQ and ED resonance wavelengths respectively (the TMOKE case). As shown in figure 4a and 4c, for the MD/MQ modes, the resonances show characteristic circular electric field distribution. Thus the $E_z$ field is significantly enhanced, inducing circular displacement currents in Ce:YIG. Whereas for the ED resonances shown in Fig. 4b and 4d, the electric dipole resonance induces mostly $E_x$ or $E_y$ fields in Ce:YIG due to the oscillated dipole resonance behavior. Considering the magneto-optical effect, only the electric field perpendicular to the applied magnetic field shows the magneto-optical effects, as indicated by the form of the permittivity tensors in Eq. (1). Therefore, $E_z$ makes a main contribution for the near normal incident LMOKE-T and s-polarized TMOKE. We can quantitatively compare the $E_x$, $E_y$, $E_z$ field intensity by performing an area integral of |E| inside the Ce:YIG layer. The integrated $E_z$ intensity is 1.34 times higher than $E_x$ at the MD-WG mode wavelength, whereas the intensity of $E_x$ is 4.6 times larger than $E_z$ at the ED-WG resonance mode for $\theta=3°$. For $\theta=45°$ incident angle of the s-polarized TMOKE configurations, the intensity of $E_z$ is comparable with $E_y$ at the MQ mode wavelength, whereas the intensity of $E_y$ is 1.87 times larger than $E_z$ at the ED resonance. Fig. 4e and 4f show the relationship between the incident wavevector, the electric field vector and the applied magnetic field directions. For the resonances dominated by MD/MQ, the p-polarized/s-polarized incident light generates $E_z$ component of the displacement current, leading to enhancement of Kerr effect when the magnetic field is along the x (LMOKE-T configuration) or y directions (TMOKE configuration). On the contrary, the electric field distribution for ED resonance modes is different. The electric fields remain mostly along x- or y-directions. Nevertheless, a small amount of $E_z$ field is also observed for these modes, therefore we do see TMOKE/LMOKE-T enhancement at around ED resonances but with a much lower amplitude compared to MD/MQ modes. These observations again highlight the important role of circular displacement currents to the observed anomalous magneto-optical effects.

The observation of several anomalous magneto-optical effects in high index contrast, all-dielectric magneto-optical metasurfaces indicate unprecedented opportunity of using the complex modal profiles in high-index-contrast all-dielectric Mie resonators to control light propagation by magnetization and vice versa. Study on other dielectric resonance modes can be envisioned for future works, such as anapole modes[39], Fano resonance modes[40], supercavity modes[41] as well as coupled Mie resonance modes[42]. Ultra-high quality factor modes such as the bound states in the continuum modes[43] can also be explored. Note the observed magneto-optical effects are rooted in *localized* Mie resonance modes, which is very different from several previous proposals of enhancing magneto-optical effects by propagating waveguide modes[29, 30, 44]. This new mechanism offers a possibility to construct advanced magneto-optical materials by locally design the structure at the subwavelength scale, leading to a variety of possibilities to control the wave front by specifically designed magneto-optical Mie resonators and metasurfaces. On the other hand, the complex field profile also indicates rich physics of manipulating spin using femtosecond optical pulses, *i.e.,* ultrafast optomagnetic effects in high index contrast Mie resonators[45]. The low optical absorption, strong field localization, broad angular and frequency width and designable modal profiles may enable more efficient all optical magnetization switch for future spintronic devices. These possibilities make high index contrast, all-dielectric magneto-optical metasurfaces promising for a variety of applications including vectoral magnetic field sensing, free space non-reciprocal photonic devices, magneto-optical imaging and optomagnetic memories.

## CONCLUSIONS

In summary, we observe anomalous magneto-optical effects including giant s-polarized TMOKE and LMOKE-T in high index contrast magneto-optical Mie resonators, which are not achievable in bulk or planar magneto-optical materials. These magneto-optical effects are originated from the unique circular displacement currents associated with MD or MQ modes in high index contrast all-dielectric Mie resonators. A giant s-polarized TMOKE up to 6.4 % and nearly two orders of magnitude enhancement of the LMOKE-T under near normal incidence conditions are observed experimentally. Our results indicate the possibility of utilizing the complex Mie resonance modes to realize novel magneto-optical effects, which will allow unprecedented opportunity to control light propagation with magnetization and vice versa. These new observations are promising for a variety of applications including vectoral magnetic field sensing, free space non-reciprocal photonic devices, magneto-optical imaging and optomagnetic memories.

## METHODS

**Numerical Simulation** Numerical simulations were carried out using finite element method via COMSOL MUITIPHYSICS. Periodic boundary conditions were applied in both x and y directions for the unit cell. Perfect Match Layer (PML) were set along the z direction at the upper and lower boundaries of the structure. The refractive index of the quartz substrate was set as 1.46 in our model. As for the permittivity tensor of Ce:YIG films, we adopted the detailed parameters from reference 24. The mesh sizes varied slightly depending on the refractive index of the material to ensure the accuracy of the results. The transmittance and reflectivity of the metasurfaces were obtained by extracting the S parameters of the scattering matrix. Then the TMOKE and LMOKE-T angle can be calculated using Eq. (3) and Eq. (4), respectively. The RCWA method was also applied for angle dependent transmission spectrum simulations. And the calculated results agree well with numerical simulation by finite element analysis using COMSOL (seen in Supplementary Fig. S4).

**Calculation of Waveguide Mode Dispersion** To further understand the WG modes, we use the equivalent medium method to treat the structure of 'Si nanodisks + magneto-optical film' as an equivalent slab waveguide core surrounded by air and SiO2 claddings. According to the planar waveguide theory[18, 19, 26], the dispersion equation of the modes can be defined as:

$$(\frac{2\pi}{\lambda}\sin\theta + \frac{2\pi}{P}m_x)^2 + (\frac{2\pi}{P}m_y)^2 = (\frac{2\pi}{\lambda}n_\beta)^2 \quad (4)$$

where $\lambda$ is wavelength of incident light from the free space, $\theta$ is the angle of incidence, $m_{x/y}$ is the diffraction orders along x and y directions, $P$ is the period of the Si nanodisks, $n_\beta$ is the effective refractive index of the guided mode, $2\pi n_\beta/\lambda$ is the wave vector of the guided mode. Firstly, we assume that the resonance at 1310 nm wavelength under normal incidence is a waveguide mode, then we can obtain the $n_\beta$ by equation $n_\beta = \frac{\lambda}{P}\sqrt{(m_x^2 + m_y^2)}$. With this fixed $n_\beta$, the trend of waveguide mode as a function of angles and wavelengths can be calculated via

$$\sin\theta = \pm\sqrt{n_\beta^2 - (\frac{m_y\lambda}{P})^2} - \frac{m_x\lambda}{P}, \text{ as shown in Fig. 1e and 1f.}$$

**Multipole Decomposition** By extracting the value of electric field E(r) through simulation calculation, we defined the induced polarization current density $\mathbf{J}(\mathbf{r}) = -i\omega\varepsilon_o[\varepsilon_r(\mathbf{r})-1]\mathbf{E}(\mathbf{r})$, then the electric $a(l,m)$ and magnetic $b(l,m)$ spherical multiple coefficients can be calculated as[46]:

$$a(l,m) = \frac{(-i)^{l-1}k\eta}{2\pi E_0} \frac{\sqrt{(l-m)!}}{\sqrt{l(l+1)(l+m)!}} \int \exp(-im\phi) \left\{ \begin{array}{l} [\psi_l(kr)+\psi''_l(kr)]P_l^m(\cos\theta)\hat{\mathbf{r}}\cdot\mathbf{J}(\mathbf{r}) + \\ \frac{\psi'_l(kr)}{kr}\left[\frac{d}{d\theta}P_l^m(\cos\theta)\hat{\theta}\cdot\mathbf{J}(\mathbf{r}) - \frac{im}{\sin\theta}P_l^m(\cos\theta)\hat{\phi}\cdot\mathbf{J}(\mathbf{r})\right] \end{array} \right\} d^3\mathbf{r} \quad (5)$$

$$b(l,m) = \frac{(-i)^{l+1}k^2\eta}{2\pi E_0} \frac{\sqrt{(l-m)!}}{\sqrt{l(l+1)(l+m)!}} \int \exp(-im\phi) j_l(kr) \left[\frac{im}{\sin\theta}P_l^m(\cos\theta)\hat{\theta}\cdot\mathbf{J}(\mathbf{r}) + \frac{d}{d\theta}P_l^m(\cos\theta)\hat{\phi}\cdot\mathbf{J}(\mathbf{r})\right] \quad (6)$$

where $E_0$ is the electric field amplitude of the incident plane wave; $\eta$ is the impedance of free space; $j_l(kr)$ is the spherical Bessel function of the first kind while $P_l^m(\cos\theta)$ is the associated Legendre polynomials. Then the total contribution to the scattering cross section $C_{scat}$ of the Si disk can now be written as

$$C_{scat} = \frac{\pi}{k^2} \sum_{l=1}^{\infty} \sum_{m=-l}^{l} (2l+1)(|a(l,m)|^2 + |b(l,m)|^2) \tag{7}$$

For the sake of simplicity, we investigate the scattering response of a single silicon disk in homogeneous medium with the refractive index of 1.46.

**Sample Fabrication** The magnetic oxides were deposited on 10 mm × 10 mm double-polished quartz substrate by pulsed laser deposition (TSST PLD, Netherlands) equipped with 248 nm KrF excimer laser. A layer of 50 nm YIG was first deposited on $SiO_2$ at room temperature with the oxygen pressure of 5 mTorr. Then a rapid thermal annealing process in oxygen atmosphere of 2 Torr at 900 °C for 480 s was performed to ensure crystallization. Subsequently, a 200 nm thick Ce:YIG layer was deposited at 750 °C with the oxygen pressure of 10 mTorr using 2.03 $J/cm^2$ power density. After deposition, the film was cooled down in the main chamber at the rate of 5 °C/min. The XRD pattern and magneto-optical response of the oxide film grown on quartz can be seen in Fig. S1. The α-Si nanodisk arrays were fabricated by electron beam lithography (EBL, Raith). First, a layer of uniform amorphous silicon was deposited by plasma enhanced chemical vapor deposition (PECVD). Then the pillar-shaped patterns with an area of 200 μm × 200 μm was prepared by electron-beam lithography using HSQ resist (XR-1541006) followed by deep reactive ion etching (DRIE). The etching gas and specific flow ratio applied in the experiment were $CH_3F_3$: $SF_6$: $O_2$ = 30: 30: 5.

**Optical and Magneto-optical Measurement** The transmittance spectra and LMOKE-T effects of the metasurfaces were measured by a custom-built characterization set-up as shown in Fig. S8. The incident light was provided by a supercontinuum laser (NKT Photonics) connected with a spectrometer. Then the beam was incident on the sample surface through an aperture and a Glan-Taylor calcite polarizer to ensure linear polarization in the x direction. A pair of lenses of the same focal length were equipped on both sides of the sample to achieve confocal effect. The light spot can be focused into the size of 10 μm to meet the test requirements. LMOKE-T spectra were obtained with the applied magnetic field along x-direction. The magnetic field is generated by an electromagnet with the maximum magnetic field of 1T. The reflection and TMOKE spectra were characterized on a spectroscopic ellipsometer (J. A. Woollam RC2). The reflectance of the silicon wafer with a 25 nm silicon dioxide layer was firstly measured, then we used this as the baseline to obtain the reflection spectra of the structure at the 45˚ incident angle. The applied magnetic field of 3 kOe is along the in-plane y direction, which was provided by a neodymium iron boron permanent magnet. By changing the position of the permanent magnet, the reflection spectra of the structure under the positive and negative magnetic field were measured, then the spectra of TMOKE were calculated according to equation 3. As for the TMOKE hysteresis, it was also measured by the custom-built characterization set-up. The magnetic field generated by an electromagnet is applied in y-direction and the reflectivity of the sample changed with the applied magnetic field was measured. Then the hysteresis was obtained via using the equation $\delta=\frac{R(M)-R(0)}{R(0)}$ [47], where R(0) is the reflectivity through the non-magnetized sample.

## DATA AVAILABILITY

All the data generated and analyzed during this study are included in the article and its Supplementary Information. Source data are provided with this paper.

## ACKNOWLEDGEMENTS

The authors are grateful for support by the National Natural Science Foundation of China (NSFC) (Grant Nos. 51972044 and 52021001), Ministry of Science and Technology of the People's Republic of China (MOST) (Grant Nos. 2016YFA0300802 and 2018YFE0109200), Sichuan Provincial Science and Technology Department (Grant Nos. 2019YFH0154 and 2020ZYD015), the Open-Foundation of Key Laboratory of Laser Device Technology, China North Industries Group Corporation Limited (Grant No. 200900), and the Fundamental Research Funds for the Central Universities (Grant No. ZYGX2020J005). D.O.I. and V.I.B. acknowledges financial support by the Ministry of Science and Higher Education of the Russian Federation, Megagrant project N 075-15-2019-1934.

## AUTHOR CONTRIBUTIONS

S.X., J.Q., L.D. and L.B. designed the experiments. Q.L., Y.C. and H.D. fabricated the magnetic all-dielectric metasurface samples. S.X. and W.Y. deposited the magneto-optical films and measured the transmittance, reflectance and magneto-optical spectra. H. W. and Y. L. helped with the material and device structure characterizations. D.O.I. and V.I.B. calculated the angle-resolved transmittance spectra using RCWA and the dispersion of the WG modes. O.N. and M.V. helped the theoretical analysis and the set-up of the home-made magneto-optical spectrum characterization stage. All the authors contributed to the preparation of the manuscript.

## COMPETING INTERESTS

The authors declare no competing interests.

## ADDITIONAL INFORMATION

**Supplementary Information**

**Correspondence and requests for materials** should be addressed to J.Q., V.I.B., Y.C., M.V. or L.B.